\documentclass[aps,pre,twocolumn,showpacs,superscriptaddress,12p,longbibliography,floatfix]{revtex4-2}

\usepackage{amsmath}
\usepackage{amssymb}
\usepackage[colorlinks=true,linkcolor=red]{hyperref}
\usepackage[sort&compress]{natbib}
\usepackage{graphicx}
\usepackage{color}
\usepackage{amsmath}
\usepackage{siunitx}
\usepackage{quantikz}

\newcommand{\RNumb}[1]{\uppercase\expandafter{\romannumeral #1\relax}}
\newcommand*\subtxt[1]{_{\textnormal{#1}}}

\newcommand{\new}[1]{\textcolor{black}{#1}}

\begin{document}

\title{Tunable coupling scheme for implementing two-qubit gates on fluxonium qubits}

\author{I. N. Moskalenko}
\affiliation{National University of Science and Technology "MISIS", 119049 Moscow, Russia}
\affiliation{Russian Quantum Center, 143025 Skolkovo, Moscow, Russia}
\author{I. S. Besedin}
\email[Corresponding author: ]{ilia.besedin@gmail.com}
\affiliation{National University of Science and Technology "MISIS", 119049 Moscow, Russia}
\affiliation{Russian Quantum Center, 143025 Skolkovo, Moscow, Russia}
\author{I. A. Simakov}
\affiliation{National University of Science and Technology "MISIS", 119049 Moscow, Russia}
\affiliation{Russian Quantum Center, 143025 Skolkovo, Moscow, Russia}
\affiliation{Skolkovo Institute of Science and Technology, 143026 Moscow, Russia}
\affiliation{Moscow Institute of Physics and Technology, 141701 Dolgoprundy, Russia}
\author{A. V. Ustinov}
\affiliation{National University of Science and Technology "MISIS", 119049 Moscow, Russia}
\affiliation{Russian Quantum Center, 143025 Skolkovo, Moscow, Russia}
\affiliation{Physikalisches Institut, Karlsruhe Institute of Technology, Karlsruhe, Germany}

\date{\today}

\begin{abstract}
	The superconducting fluxonium circuit is an RF-SQUID-type flux qubit that uses a large inductance built from an array of Josephson junctions or a high kinetic inductance material. This inductance suppresses charge sensitivity exponentially and flux sensitivity quadratically. In contrast to the transmon qubit, the anharmonicity of fluxonium can be large and positive, allowing for better separation between the low energy qubit manifold of the circuit and higher-lying excited states. Here, we propose a tunable coupling scheme for implementing two-qubit gates on fixed-frequency fluxonium qubits, biased at half flux quantum. In this system, both qubits and coupler are coupled capacitively and implemented as fluxonium circuits with an additional harmonic mode. We investigate the performance of the scheme by simulating a universal two-qubit fSim gate. 
	In the proposed approach, we rely on a planar on-chip architecture for the whole device. Our design is compatible with existing hardware for transmon-based devices, with the additional advantage of lower qubit frequency facilitating high-precision gating.
		 
\end{abstract}
\maketitle


Quantum superconducting circuits based on Josephson tunnel junctions are a flexible platform for building artificial atoms. Rapid progress has been made in the last decade due to appearance of new types of qubits \cite{Oliver, Xiang} and improvements in coherence properties \cite{Houck}. Successful prototypes of superconducting quantum processors developed by different research groups \cite{volume, Google, Rigetti} to date are based on transmons, which have shown the best gate fidelities among superconducting qubits. Despite the relatively high values of coherence times of transmons in the order $100\ \si{\micro\s}$ they are outperformed by an order magnitude in $T_1$ coherence times by fluxonium qubits \new{\cite{Fluxonium, Pop}}.
The spectra of transmon qubits are similar to those of weakly anharmonic oscillators. Although multiqubit processors with efficient two-qubit
gates\cite{volume, Google, Rigetti} have already been demonstrated, weak anharmonicity of their base elements presents a significant challenge for further scaling them up and improving gate fidelities.

A changeover to fluxonium qubits could provide a possible upgrade path towards large-scale superconducting quantum processors \cite{Fluxonium, Manucharyan, Nguyen, Masluk, Pop, heavy-fluxonium} as fluxoniums have millisecond energy relaxation times at flux degeneracy point. Such long lifetime of the first excited state is partially due to its very low (hundreds of megahertz) transition frequency from the ground state. \new{This leads to lower decay rates, since single-photon dielectric loss tangents only weakly depend on frequency\cite{Skacel2015}}. Low transition frequencies, however, lead to operation of the qubit in a relatively ``hot'' environment. Because of this, qubits can't be initialized in the ground state by passive thermalization. However, in a practical quantum processor qubit state initialization can be realized by fast active reset \cite{Reset}.
Promising coherence times \new{($>200\ \si{\micro\s}$)} have already been obtained in \new{chip-integrated fluxoniums  \cite{Zhang2021}, while in 3D cavities coherence times exceed even 1 ms \cite{ms}.} In a recent work \cite{CZfluxoniums} first microwave-activated CZ gates have been demonstrated also in a 3D cavity. \new{Recently, another type of  microwave-activated two-qubit gate has been proposed, the bSWAP gate \cite{Nesterov2021}.}
However, high-fidelity two-qubit gates in planar geometry are yet to be demonstrated. Moreover, scaling up beyond two qubits is extremely challenging in a 3D architecture.

\begin{figure}
	\includegraphics[width=0.95\columnwidth]{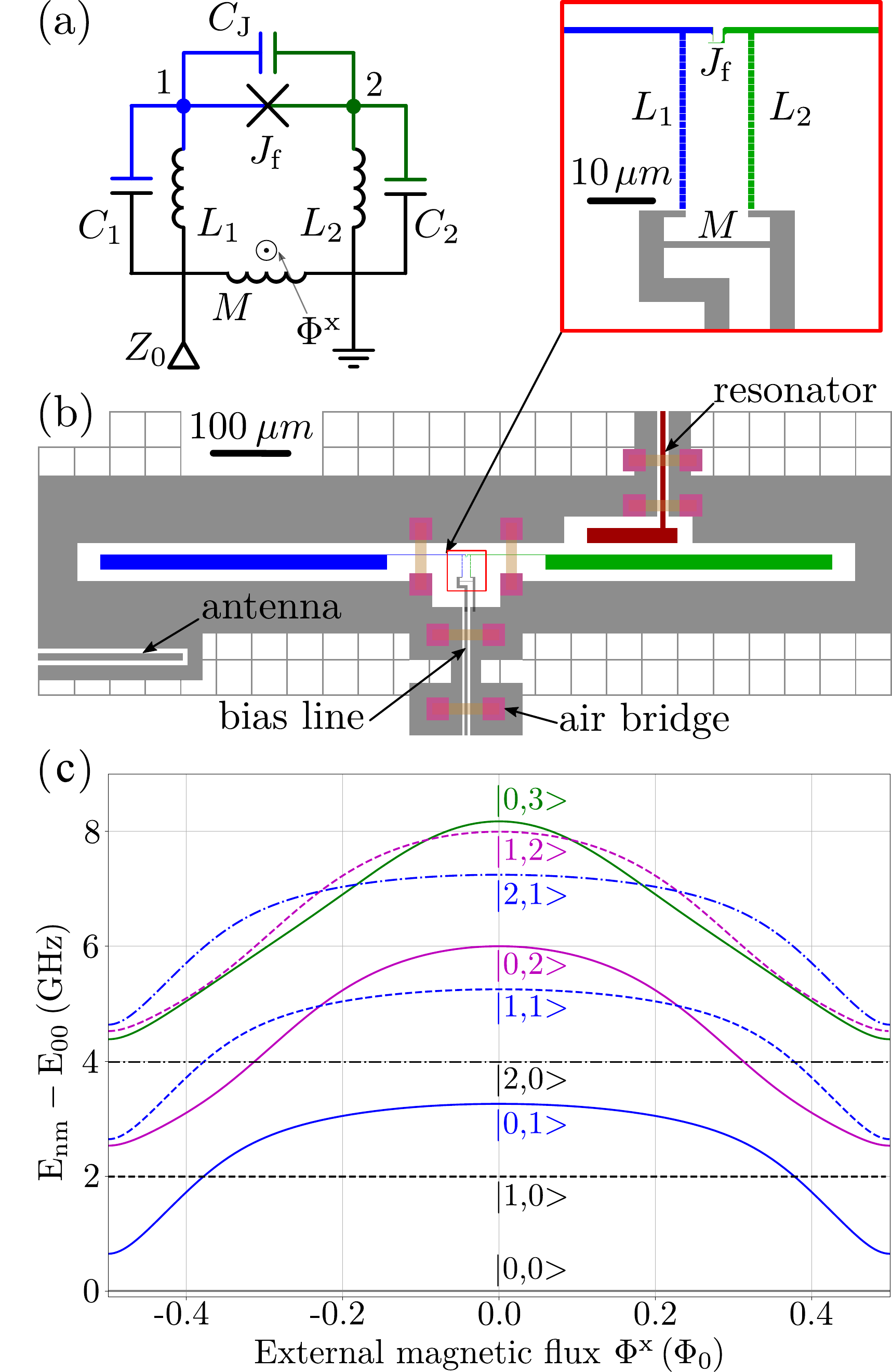}
	\caption{(color online) (a) Modified fluxonium circuit diagram, consisting of one Josephson junction, two large inductors and three capacitors. (b) Concept layout with readout resonator and bias line for magnetic flux control. (c) Energy levels of the modified fluxonium system vs external magnetic flux ${\Phi^{\textnormal{x}}}$ for $E_\textnormal{J}=2.24\ \si{\giga\hertz}$, $E_\textnormal{L}=1.64\ \si{\giga\hertz}$, $C_{1,2}=70.1\ \si{\femto\farad}$,  $C_\textnormal{J}=1.3\ \si{\femto\farad}$}
	\label{fig:new_fluxonium}
\end{figure}

\new{In this work, we consider a specific parameter regime of fluxonium which allows strong capacitive coupling to the qubit transition. In terms of frequency and anharmonicity it is close to conventional fluxonium \cite{Fluxonium, Manucharyan, Nguyen}, while the ratio between the Josephson and shunt inductance is close to the quarton regime \cite{G-flux}. At the same time, the charging energy is relatively high: $E_J \sim E_L \sim 4E_C$. A detailed comparison is given in the Supplementary Information.} The circuit consists of two superconducting islands connected with a small Josephson junction, and inductively shunted to the ground electrode (Fig.~\ref{fig:new_fluxonium}a).
The proposed fluxonium can be utilized as the unit cell (both qubit and coupler) for a scalable quantum processor. A possible layout corresponding to realistic capacitances and inductances is shown in Fig.~\ref{fig:new_fluxonium}b.
Neighboring qubits can be capacitively coupled, allowing to adapt the simple and broadly applicable capacitive tunable coupling scheme \cite{Oliver1, TunableCoupler, Google}.

The scheme that we propose here consists of two fluxonium qubits with a tunable coupler between them, which by itself is also a fluxonium qubit. Both computational qubits are biased at the flux degeneracy point. The interaction strength between the qubits is controlled by the central ``coupler'' fluxonium flux bias. At the flux degeneracy point, all three qubits are close to resonance and exhibit a strong $XX$-type interaction. Away from it, only a small residual $ZZ$-type interaction between the qubits is left. \new{A $\sqrt{\mathrm{iSWAP}}$-like gate is performed by tuning the coupler from the upper flux sweet stop to the lower sweet spot, waiting quarter of} a vacuum Rabi cycle, and tuning back. Using numerical simulation, we demonstrate how decoherence, leakage and coherent errors can affect the gate performance.

The proposed scheme is compatible with existing hardware, moreover, the additional advantage of this approach is the ability to use lower frequency electronics for qubit and coupler control. Switching to sub-gigahertz controls could drastically reduce the cost and complexity of the control electronics and wiring.


A modified fluxonium circuit and a possible layout are shown in Fig.~\ref{fig:new_fluxonium}. It consists of a Josephson junction with energy $E_{\textnormal{J}}$ shunted by a capacitance $C_\textnormal{J}$ and two large (super-) inductors $L_{1}$ and $L_{2}$ linked to form a loop.  Superinductances $L_{1,2}$ can be built from long arrays ($>50$) of large identical Josephson junctions.
Both nodes $1;2$ have a distributed mutual capacitance with the ground node $C_{1;2}$. \new{External magnetic flux $\Phi^\mathrm{x}$ can be applied with a current bias line, which is grounded through a part of the fluxonium loop. The inductance of that wire $M$ determines how much current is required to tune the qubit frequency from maximum to minimum. We neglect the influence of this inductance for the qubit Hamiltonian, as it is several orders of magnitude smaller that the large inductances $L_1$ and $L_2$.}

The circuit has two degrees of freedom. We denote the nodal phases as $\varphi_1$ and $\varphi_2$. Due to the circuit's symmetry, the normal mode coordinates of the circuit are defined as:
\begin{multline}
\vartheta^+ = \varphi_1 + \varphi_2; \ \ \ \ \vartheta^- = \varphi_1 - \varphi_2. \ \ \ \ \ \ \ \ \ \    
\label{eq1} 
\end{multline} 

The $\vartheta^-$-mode is associated with a phase differences across the Josephson junction and is thus nonlinear, the $\vartheta^+$-mode does not bias the junction and is therefore a fully harmonic mode. In the absence of disorder among circuit elements $L_1=L_2=L$, $C_1=C_2=C$ \new{the modes are decoupled, and the Hamiltonian is}
\begin{multline}
\hat{H} = \hat{H}\subtxt{h} + \hat{H}\subtxt{f}, \ \ \ \ \ \ \ \ \ \ \ \ \ \
\label{eq2}
\end{multline}
\begin{multline}
\hat{H}\subtxt{h} =  4E_{\textnormal{Ch}}(\hat{n}^+)^2 + \frac{1}{2}E_\textnormal{L}(\hat{\vartheta}^+-{\varphi}^x)^2 , \ \ \ \ \ \ \ \
\label{eq3}
\end{multline}
\begin{multline}
\hat{H}\subtxt{f} = 4E_{\textnormal{Cf}}(\hat{n}^-)^2  + \frac{1}{2}E\subtxt{L}(\hat{\vartheta}^--{\varphi}^x)^2 + 
E_{\textnormal{J}}[1-\cos(\hat{\vartheta^-})],
\label{eq4}
\end{multline}
where $\hat{n}^-$ and $\hat{n}^+$ are the canonically conjugate Cooper pair numbers to $\hat{\vartheta}^-$ and $\hat{\vartheta}^+$, respectively. Here we also introduce a dimensionless variable for external flux $\varphi^\textnormal{x}=\frac{2\pi{\Phi}^\textnormal{x}}{\Phi_0}$, and convert the circuit element parameters to energy units $E\subtxt{L}=(\Phi_0/2\pi)^2/2L$, $E_{\textnormal{Cf}} = e^2/2C\subtxt{f}$, where $C\subtxt{f}=(C+C\subtxt{J})/2$, $E_{\textnormal{Ch}} = e^2/2C\subtxt{h}$, where $C\subtxt{h}=C/2$.
 
Mutual capacitance between the fluxonium mode and other circuit elements is a scarce resource. Increasing the absolute value of a mutual capacitance also increases the total capacitance of the fluxonium mode, which drives down the qubit frequency and decreases the coupling strength of the fluxonium to everything else. 
This contrasts with inductively coupled fluxonium qubits, where the coupling strength does not directly depend on the qubit frequency.
The two-island configuration of the fluxonium qubit can utilize either of the two islands to couple to other elements, while the total effective capacitance is half of the total capacitance of each of the islands relative to the ground electrode. This configuration allows us to work in the $300-700\ \si{\mega\hertz}$ qubit frequency range at the operating point and still have large coupling strengths between neighboring fluxoniums.

The computed energy spectrum for our qubit as a function of external flux \new{$\Phi^\textnormal{x}$} is plotted in Fig.~\ref{fig:new_fluxonium} (c). The circuit parameters are $E\subtxt{J}=2.24\ \si{\giga\hertz}$, $E\subtxt{L}=1.64\ \si{\giga\hertz}$, \new{$C=70.1\ \si{\femto\farad}$},  $C_\textnormal{J}=1.3\ \si{\femto\farad}$. These circuit parameters will be further used for the tunable coupler. The eigenstates are labeled as $\ket{n\subtxt{h}, n\subtxt{f}}$, where $n\subtxt{h}$ is the harmonic mode occupancy and $n\subtxt{f}$ is the fluxonium mode occupancy. The harmonic mode frequency is $2.0\ \si{\giga\hertz}$. The fluxonium mode fundamental transition frequency $f_\textnormal{Q}$ spans from $625\ \si{\mega\hertz}$ at the flux degeneracy point to $3.31\ \si{\giga\hertz}$ at zero flux bias. The fluxonium mode anharmonicity $\delta f_\textnormal{Q}$ at the flux degeneracy point is around $1.911\ \si{\giga\hertz}$.  

\new{The flux bias line is coupled to the fluxonium mode of the qubit, allowing to perform both excitation and qubit frequency control with a single wire. This approach has been used to reduce wiring complexity in large NISQ devices \cite{Arute2019}. However, if the inductance $M$ is too large, it becomes a significant decay channel for the qubit excitation. The decay rate of this process can be obtained through Fermi's Golden rule:
\begin{equation}
    \gamma = \omega \frac{R_Q}{2Z_0}\left(\frac{M}{L_1+L_2}\right)^2 \left|\langle 0 |\hat{\vartheta^{-}}|1\rangle\right|^2,
\end{equation}
where $\omega$ is the qubit frequency, $Z_0 = \SI{50}{\ohm}$ is the control line impedance, $R_Q$ is the von Klitzing constant, and $\langle 0 |\hat{\vartheta^{-}}|1\rangle$ is the matrix element of the fluxonium mode phase operator for the fundamental transition.
We choose $M=12~\mathrm{pH}$ for the control wire inductance, which corresponds to a relaxation time of 1 ms in the flux degeneracy point. Inducing half a flux quantum in the SQUID loop requires \SI{83}{\micro \ampere} of current. Due to the lower frequency of the fluxonium, this current is lower than the current required to induce the same flux in the SQUID of a typical transmon with the same decay rate into the flux line. Lower control signal amplitudes are beneficial because they help reducing RF crosstalk and give more flexibility in signal chain attenuation and filtering at low temperatures.}

\begin{figure}
	\includegraphics[width=1\columnwidth]{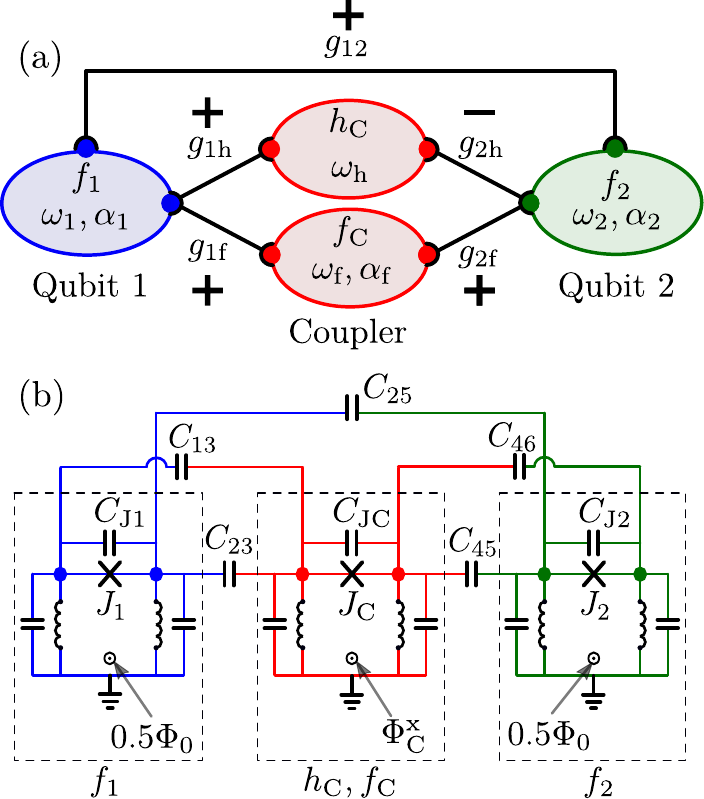}
	\caption{(color online) (a) Simplified system schematic. Two fluxonium qubits ($f_{1;2}$) are capacitively coupled via a coupler with harmonic ($h\subtxt{C}$) and tunable fluxonium ($f\subtxt{C}$) modes. \new{The plus and minus signs denote the sign of the $XX$ coupling constant between the corresponding modes. }(b) Electric circuit schematic. Each mode is highlighted in different colours (qubit mode 1 (blue), qubit mode 2 (green), and coupler mode c (red)). The computational qubits are biased \new{at} the flux degeneracy point.}
	\label{fig:Conceptual_schematic}
\end{figure}

A simplified scheme of the \new{two qubit coupling} design is shown in Fig.~\ref{fig:Conceptual_schematic}(a). The system has three qubit-qubit coupling channels: direct capacitive coupling, fluxonium mode-mediated coupling and harmonic mode-mediated coupling. 
Due to the different symmetries of the harmonic mode and the fluxonium mode, the coupling constants resulting from them have different signs. By carefully choosing the mutual capacitances and mode frequencies, we aim to utilize the destructive interference between the coupling channels and minimize the static ZZ interaction between the qubits near the zero flux bias point of the coupler. 

\new{The harmonic modes of the qubits also interact with the coupler. Since these modes are out of resonance with the computational subspace, we exclude them in the simulation of gate dynamics for the sake of computational efficiency. However, due to their non-negligable contribution to the crosstalks, coupling to these modes is accounted for in the calculation of static coupling terms.}

The electric circuit schematic is shown in Fig.~\ref{fig:Conceptual_schematic}b.
It consists of two computational fluxonium qubits ($f_1$, $f_2$) each coupled to a tunable coupler with fluxonium ($f\subtxt{C}$) and harmonic ($h\subtxt{C}$) modes with a coupling strength $g_{j\textnormal{f}}$ and $g_{j\textnormal{h}}$ (j = 1, 2), as well as to each other with a coupling strength $g_{12}$. The Hamiltonian for the circuit is:
\begin{multline}
	\hat{H}\subtxt{full} = \hat{H}_{\textnormal{f1}}  + \hat{H}_{\textnormal{hc}} +  \hat{H}_{\textnormal{fc}} + \hat{H}_{\textnormal{f2}}  + \hat{H}\subtxt{V}
	\label{eq5}
\end{multline}

where first four terms describe the independent Hamiltonians for qubit and coupler modes and $\hat{H}_{V}$ is responsible for the effective qubit-qubit interaction. The interaction term has five contributions (see Supplementary Information for the derivation): one term due to direct qubit-qubit coupling (capacitive connection between the blue and green nodes), and four terms corresponding to the interaction of either of the qubits to either of the coupler modes (capacitive connection to red nodes in Fig.~\ref{fig:Conceptual_schematic}b).\new{ Due to the different symmetries of the harmonic and fluxonium modes of the coupler, effective couplings mediated by these modes interfere destructively, allowing to cancel out either the XX or the ZZ coupling completely \cite{Mundada2019}. }

The natural gate available for this device is an iSWAP-like fSim gate\cite{fSimGate}. In our simulation, the gate is executed by applying a time-dependent flux to the coupler, changing the coupler's fluxonium mode frequency $f\subtxt{C}$. As the coupler's fluxonium mode frequency gets close to the qubit frequencies, the mediated interaction becomes resonant and energy exchange occurs. Due to the finite anharmonicity of the fluxonium qubits, the interaction is not purely transverse.

The effective interaction strength between the qubits can be obtained by diagonalizing the full system Hamiltonian, eliminating the coupler degrees of freedom, and building an effective low-energy Hamiltonian:
\begin{multline}
	\hat{H}_\textnormal{eff}/\hbar = -\frac{1}{2}\omega_1\sigma^\textnormal{z}_1 - \frac{1}{2}\omega_2\sigma^\textnormal{z}_2 + g_\textnormal{xx}\sigma^\textnormal{x}_1\sigma^\textnormal{x}_2 + \frac{1}{4}\zeta_\textnormal{zz}\sigma^\textnormal{z}_1\sigma^\textnormal{z}_2.
	\label{eq6}
\end{multline}

\begin{figure}
	\includegraphics[width=1\columnwidth]{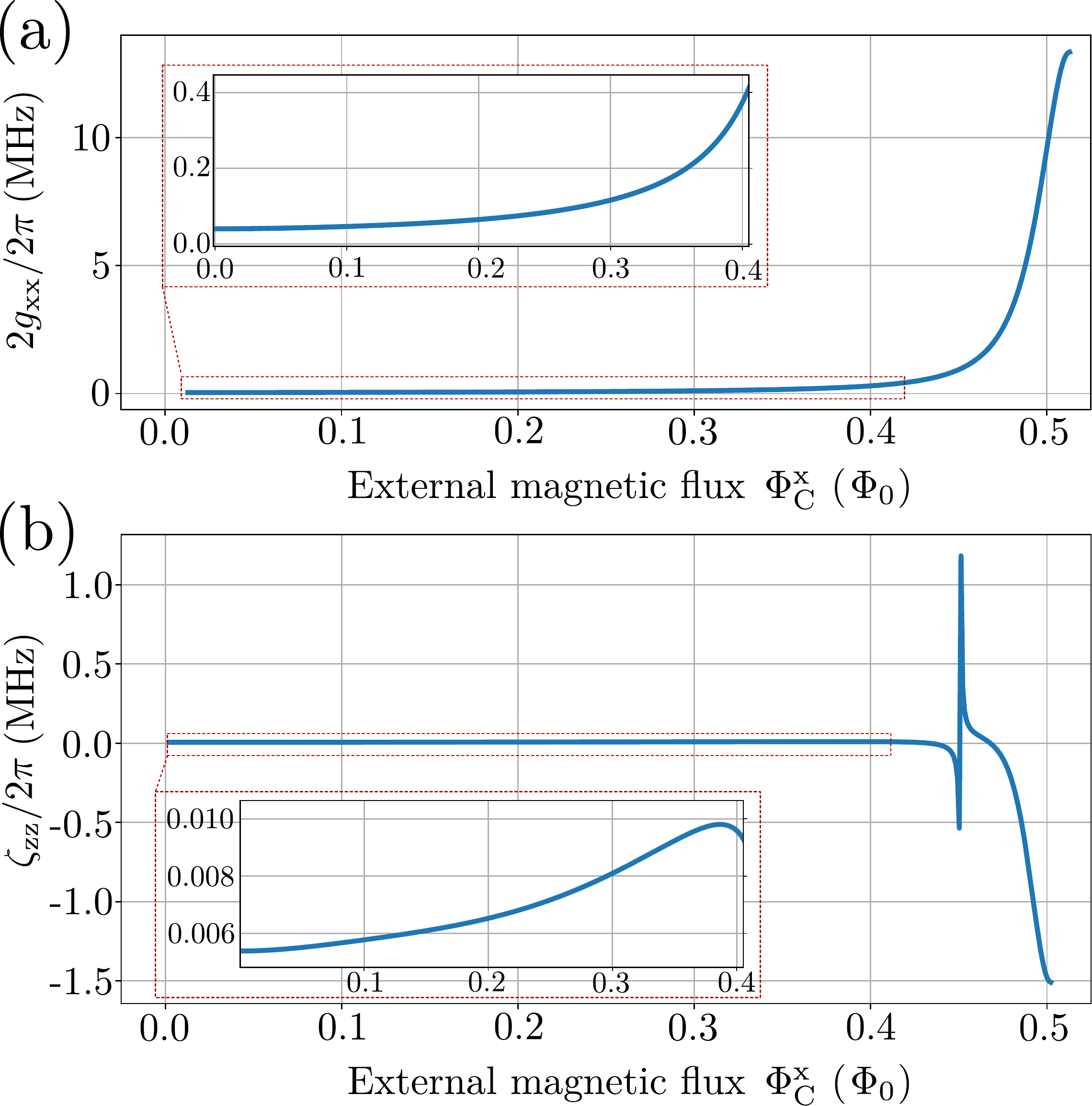}
	\caption{(color online) \new{Effective couplings as a functions of the magnetic flux threading the coupler loop}. (a) Effective transverse coupling strength  $2g_\textnormal{XX}(\Phi^\textnormal{x}_\textnormal{C})$. (b) ZZ crosstalk $\zeta_\textnormal{ZZ}(\Phi^\textnormal{x}_\textnormal{C})$}
	\label{fig:first_states_coupling}
\end{figure}

Details of the numerical calculations are presented in the Supplementary Information. 
For equal-frequency data qubits, the energy gap between symmetric and antisymmetric modes corresponds to the effective coupling $2g_\textnormal{xx}(\Phi^\textnormal{x}_\textnormal{C})$ (Fig.~\ref{fig:first_states_coupling}a). The parasitic ZZ crosstalk between $f_1$ and $f_2$ (Fig.~\ref{fig:first_states_coupling}b) is defined as $\zeta_{ZZ} = \omega_{11} - \omega_{10} -\omega_{01}$.

\begin{figure*}
    \centering
    \includegraphics[width=\textwidth]{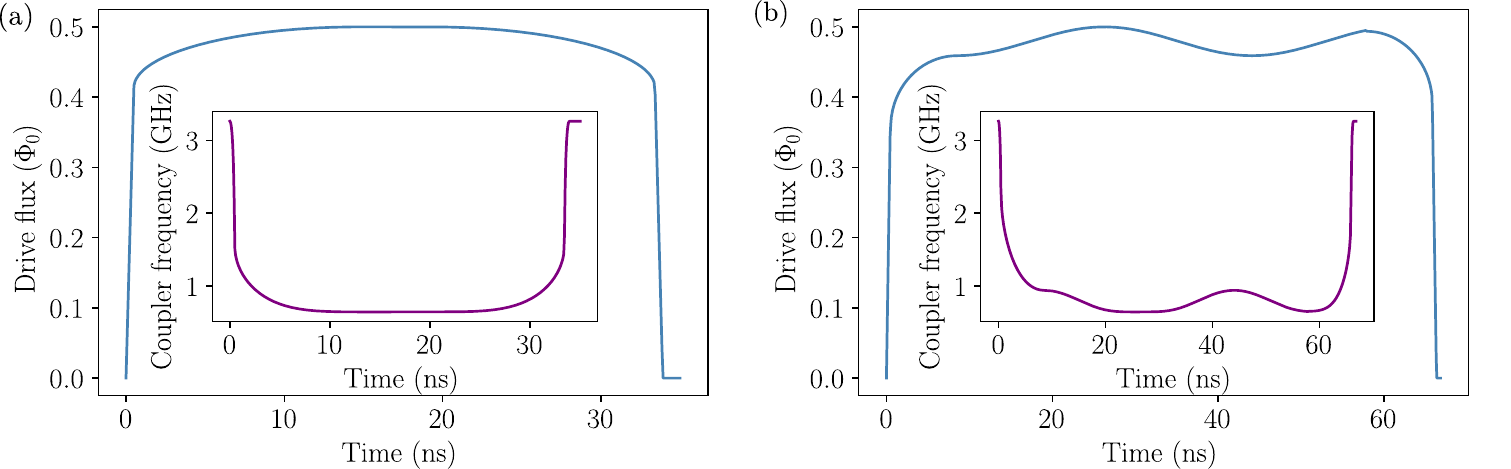}
    \caption{Shape of drive flux signal and corresponding frequency of the coupler fluxonium mode (inserted plots). (a) Data qubits have the same frequencies. The gate can be optimized over the control flux pulse rise and fall time and flat top duration. (b) Data qubits with different frequencies. Here we can also optimize the control flux pulse edges, frequency and duration of modulation.}
    \label{fig:drive}
\end{figure*}

Magnetic flux in the coupler can be used to turn on and off the effective transverse qubit-qubit interaction. Near the zero flux bias point the effective coupling is $40\ \si{\kilo\hertz}$ and increases to $13\ \si{\mega\hertz}$ at the flux degeneracy point. At the same time, the parasitic ZZ crosstalk can be reduced to around $5\ \si{\kilo\hertz}$ near the zero flux bias point. Switching between coupling on and coupling off using flux bias may induce resonant leakage into the fluxonium coupler mode, when its frequency crosses the sum of the qubit frequencies, as shown in the Supplementary Information. This resonance also gives rise in the singularity in the $\zeta_\textnormal{zz}$ dependence on flux. In the operating point ($\Phi^x_C = 0.5\Phi_0$) the parasitic ZZ crosstalk reaches \new{$\zeta_{ZZ}/2\pi=-1.5\ \si{\mega\hertz}$} and causes phase accumulation of the doubly excited state. In applications this phase accumulation can be eliminated using an echo protocol.


The fSim family of two-qubit gates \cite{Google, fSimGate} describes the set of excitation number-preserving quantum logic operations on two qubits up to single-qubit phase rotations.
Its matrix representation in the $\ket{00}$, $\ket{01}$, $\ket{10}$, $\ket{11}$ basis is given by:
\begin{equation}
    \operatorname{fSim}(\theta, \varphi)=\left(\begin{array}{cccc}
    1 & 0 & 0 & 0 \\
    0 & \cos \theta & -i \sin \theta & 0 \\
    0 & -i \sin \theta & \cos \theta & 0 \\
    0 & 0 & 0 & e^{-i \varphi}
    \end{array}\right).
    \label{eq7}
\end{equation}

Here we focus on the implementation of an \new{$\sqrt{\mathrm{iSWAP}}$}-like gate, with \new{$\theta = -\pi/4$}. Due to the non-negligible ZZ crosstalk, our gate also accumulates some small conditional phase \new{$\varphi$. An important feature of this gate is that its entangling power does not depend on $\varphi$, and two such gates can be used to construct the maximally entangling CPHASE gate (see Supplementary Material for the gate sequence). In combination with single-qubit gates, $\operatorname{fSim}\left(-\pi/4, \varphi \right)$ gates can be used to build any arbitrary two-qubit gate. }

The interaction between the computational qubits can be adiabatically turned on by slowly tuning the external magnetic flux in the coupler loop to the flux degeneracy point ($\Phi^{\textnormal{x}}_\textnormal{C} = 0.5 \Phi_0$).
Once the coupler fluxonium mode frequency is close to the frequency of \new{the} data qubits, their effective transverse coupling strength increases, inducing vacuum Rabi oscillations between them. After \new{half of a} Rabi cycle, we similarly turn off the coupler flux bias. 

The pulse should be as short as possible while remaining adiabatic with respect to leakage outside the computational subspace. The most probable leakage scenarios involve populating the coupler fluxonium mode. To avoid these transitions, \new{we use a smooth pulse shape $\Phi_\mathrm{C}^\mathrm{x}(t)$ with slow ramp close to the flux sweet spot.}

\begin{figure*}
    \centering
    \includegraphics[width=\textwidth]{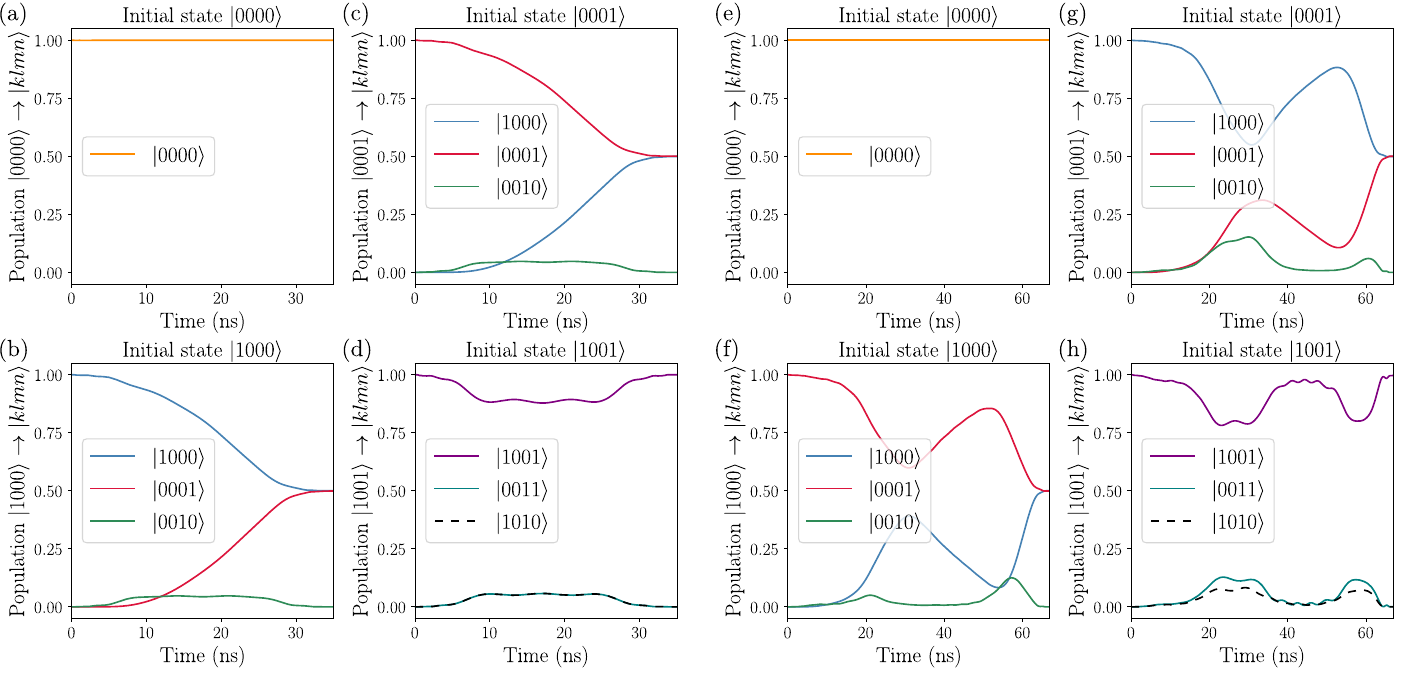}
    \caption{Time evolution of populations for four initial computational states during the gate: (a-d) qubits with the same frequency; (e-h) frequency difference of data qubits around 28 MHz. Obtained fidelities are \new{$F\approx 0.9999$ and $F\approx 0.9996$,} conditional phase $\varphi$ in \new{the fSim gate is $-0.07\pi$ and $-0.20\pi$} respectively. \new{The state notation corresponds to the mode occupations of the Hamiltonian~\eqref{eq5} as follows: $|f_1 h_C f_C f_2\rangle$, where $f_1$, $f_2$ relate to computational qubits, $h\subtxt{C}$ and $f\subtxt{C}$ are harmonic and fluxonium modes of the tunable coupler.}}
    \label{fig:populations}
\end{figure*}

The Hamiltonian of the system is given by the \new{formula~\eqref{eq5}}. In each mode of excitation, the first three energy levels are taken into account. This approximation captures the main effects of the system's evolution.
We simulate the time evolution of the system by numerically solving the Schrödinger equation with the computational stationary states as the initial conditions, and compute the projections of the resulting states onto the computational stationary states. The simulation accounts for leakage outside the computational subspace, which can occur, for example, due to excitation of the coupler degree of freedom, which results in the non-unitarity of the resulting matrix. 
To simplify further analysis, we remove the single-qubit rotations about the $z$-axis. 
We optimize the gate duration to get $\theta$ equal to $-\pi/4$. \new{The resulting 35-ns} long pulse corresponds to an fSim gate with \new{$\varphi \approx -0.07\pi$} with fidelity \new{$F \approx 0.9999$.} We use the standard expression for the two-qubit gate fidelity \cite{Fidelity}:
\begin{equation}
    F = \frac{\text{Tr}(R_{\text{ideal}}^\dagger R) + 4}{20}.
    \label{eq8}
\end{equation}

Here, $R_{\text{ideal}}$ and $R$ are Pauli transfer matrices corresponding to the actions of the closest ideal fSim gate and our simulated gate, respectively.
Time evolution of the computational states during the gate operation are presented in Fig.~\ref{fig:populations}(a-d).

In real devices, qubits may be detuned from each other. In that case, one can use a parametric modulation approach and implement the very same gate by replacing the flat-top pulse by a periodic modulation of the tunable coupler.

 Here we suggest to modulate the drive flux near the operating point ($0.5 \Phi_0$) with a sine wave profile at a frequency close to the energy difference between the fundamental transitions of the computational qubits as shown in Fig.~\ref{fig:drive}(b). In this case we get \new{$F \approx 0.9996$ with $\varphi \approx -0.20\pi$} and the dynamics of the population of the computational states is presented in Fig.~\ref{fig:populations}(e-h). In this case we have also optimized the drive pulse rise and fall times, as well as frequency and duration of the flux modulation. The entire parametric gate duration is \new{67 ns} and can be reduced further by advanced flux pulse shaping.

Finally, we perform a decoherence-aware simulation of the gate by numerically integrating the Lindblad equation with the fourth order Runge-Kutta method with different collapse operators. The gate error is calculated as $\epsilon = 1- F$ where $F$ denotes the gate fidelity, see Eq.~\eqref{eq8}.
We take into account decoherence mechanisms involving only the ground and first excited levels of each mode because the other levels are practically unoccupied during the gate time (Fig.~\ref{fig:populations}b) and hardly contribute to the resulting gate error. The collapse operators corresponding to relaxation and dephasing are defined as:
\begin{equation}
\begin{aligned}
    L_1 =\frac{1}{\sqrt{T_1}}\left(\begin{array}{ccc}
    0 & 1 & 0 \\
    0 & 0 & 0 \\
    0 & 0 & 0
    \end{array}\right) \,
    L_\varphi = \frac{1}{\sqrt{2T_\varphi}}\left(\begin{array}{ccc}
    1 & 0 & 0 \\
    0 & -1 & 0 \\
    0 & 0 & 0
    \end{array}\right)
\end{aligned}
\end{equation}
The gate errors introduced by each decoherence channel are presented in Table~\ref{error_budget}. 
\new{Apart from white noise that that can be modeled with the Lindblad equation, gates based on flux tuning of SQUIDs are susceptible to low-frequency flux noise. The characteristic time scales of this noise are usually significantly longer than the gate duration and they can be approximated by a random static flux shift during the gate. In the flux sweet spots the circuit is to first order insensitive to flux noise, leaving the rising and falling edges of the flux pulse most vulnerable to such noise. }
For the simulations we use estimates of the coherence times \new{$T_1=\SI{300}{\micro\second}$ and $T_\varphi = \SI{300}{\micro\second}$ \cite{Zhang2021}. }
In the small-error limit, errors are linear with respect to the decoherence rates. \new{Our simulation shows that the effect of decoherence on the data qubits contributes on the level of $\sim10^{-5}$ to the gate error, while the effect of coupler decoherence is by a further order of magnitude smaller.}
Taking into account the latest coherence results for fluxonium qubits in a 3D cavity \cite{Nguyen}, we believe that improvements in fabrication techniques will likely continue to enhance the coherence of planar devices. 
All time-domain simulations have been carried out using the open-source packages TensorFlow and NumPy.

\begin{table}[h!]
    \begin{center}
        \begin{tabular}{|c|c|c|c|c|c|c|c|c|c|}

        \hline
            & Unitary & \multicolumn{4}{c|}{Relaxation} & \multicolumn{4}{c|}{Dephasing} \\

            & errors &  \multicolumn{4}{c|}{\new{$T_1 = \SI{300}{\micro\second}$}} & \multicolumn{4}{c|}{\new{$T_\varphi = \SI{300}{\micro\second}$}} \\
            \cline{3-10}
            &  & $f_1$ & $f_2$ & $h_C$ & $f_C$ & $f_1$ & $f_2$ & $h_C$ & $f_C$ \\
        
        \hline 
	   	    \new{$\epsilon,\ 10^{-4}$} & \new{3.6} & \new{0.6} & \new{0.6} & \new{0.0} & \new{0.0} & \new{0.2} & \new{0.2} & \new{0.0} & \new{0.0}   \\
        \hline 

        \end{tabular}
    \end{center}
    \caption{Error budget. In the ``unitary errors'' column we show infidelity of the gate due to leakage and non-excitation-number preserving processes, and in the next eight columns we perform infidelity calculation for each decoherence channel separately.}
    \label{error_budget}
\end{table}


In conclusion, we have proposed an experimentally realizable tunable coupling scheme for implementing scalable two-qubit fSim-type gates between fluxonium qubits. The scheme is based on a simple base element with experimentally accessible circuit parameters. The performance and properties of the circuit have been simulated using numerical diagonalization of the circuit Hamiltonian.

The gate fidelity in our scheme is mainly limited by \new{unitary errors. The largest contributions to non-unitary errors come from $T_1$ and $T_\varphi$ times of the data qubits}. These coherence times have been shown to routinely exceed hundreds of microseconds in fluxonium devices.
Our proposed iSWAP-like parametrically driven gate provides a promising alternative pathway towards
high fidelity two-qubit gates using the existing transmon-based designs. We emphasize that the low frequency of fluxonium qubits opens the possibility of using sub-gigahertz wiring and electronics for gate operations.

\section*{Data availablity}

The data that supports the findings of this study are available within the article.

\begin{acknowledgments}
Development of theoretical model was supported by the Russian Science Foundation, Project (21-72-30026). Numerical simulations were supported by the Ministry of Science and Higher Education of the Russian Federation
(project no. K2A-2018-048). This work was partially supported by Rosatom.
\end{acknowledgments}


\clearpage
	
\appendix
\section{COMPARISON WITH OTHER PARAMETER REGIMES}\label{100}

\begin{figure}
	\includegraphics[width=0.9\columnwidth]{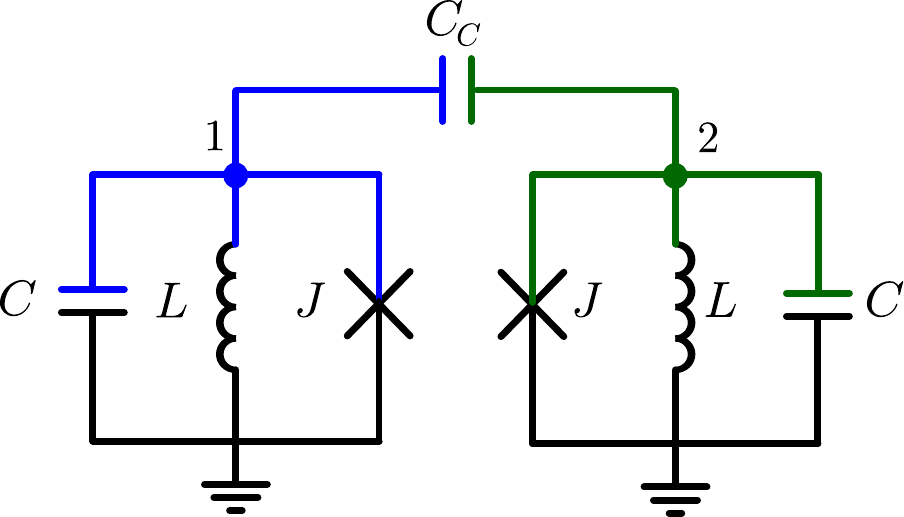}
	\caption{(color online) Equivalent lumped-element circuit for the two capacitively coupled generalized flux qubits. Each qubit circuit is highlighted in different colours (qubit 1 (blue), qubit 2 (green)). $L_i$ stand for inductors, $C_i$ stand for the capacitances with respect to the ground electrode, $C_C$ are the mutual capactitances between nodes $1$ and $2$ that facilitate coupling between the qubits.}
	\label{fig:Gflux_qubits_coupling}
\end{figure}

Fluxonium qubits can be described in the framework of the generalized flux qubit \cite{G-flux} system. For sufficiently long chains of junctions used to implement shunt inductance, generalized flux qubits are essentially RF SQUIDs and can be described by three parameters: charging energy $E_C$, Josephson energy $E_J$ and shunt inductance energy $E_L$. In compare to previous RF-SQUID type qubits, fluxonium\cite{Fluxonium} utilizes a chain of Josephson junctions, which allows to exceed the vacuum impedance and operate in the $E_J \gg E_C \gg E_C$ regime. Additional capacitive shunting of the phase slip junction and reduction of energy participation ratios of interfaces and improves coherence times\cite{Nguyen, Pop}. Extreme shunting of the phase slip junction, both inductive and capacitive, significantly lowers the qubit frequency and reduces sensitivity to AC voltage; the corresponding parameter regime has been dubbed heavy fluxonium \cite{heavy-fluxonium, heavy}. Between fluxonium-type qubits with coherent tunneling in a double-well potential and transmon qubits with plasma oscillations in a weakly anharmonic potential lies the quarton\cite{G-flux} which is characterized by $E_J=E_L$.
We compare anharmonicity, qubit frequency and coupling strength between two identical capacitively coupled qubits biased at half flux quantum for different ratios of $E_L/E_C$ and $E_J/E_C$. For this purpose we consider the Hamiltonian 

\begin{multline}
\hat{H} = \sum\limits_{\alpha=1,2}4E_{C}\hat{n}_{\alpha}^2 + E_{J}\cos\hat{\varphi}_{\alpha}+ \frac{1}{2}E_{L}\hat{\varphi}_{\alpha}^2 \\
+4\kappa E_C\hat{n}_{1}\hat{n}_{2},
\label{eqS0}
\end{multline}
which corresponds to two capacitively coupled fluxonium qubits (Fig.~\ref{fig:Gflux_qubits_coupling}). The charging energy $E_C = e^2/(2C_\Sigma)$ is defined by the effective fluxonium capacitance $C_\Sigma = (C+2C_C)/(1+C/C_Q)$, and the effective capacitive coupling ratio $\kappa = C_C/(C + C_C)$ cannot exceed 1.

\begin{figure}
	\includegraphics[width=1\columnwidth]{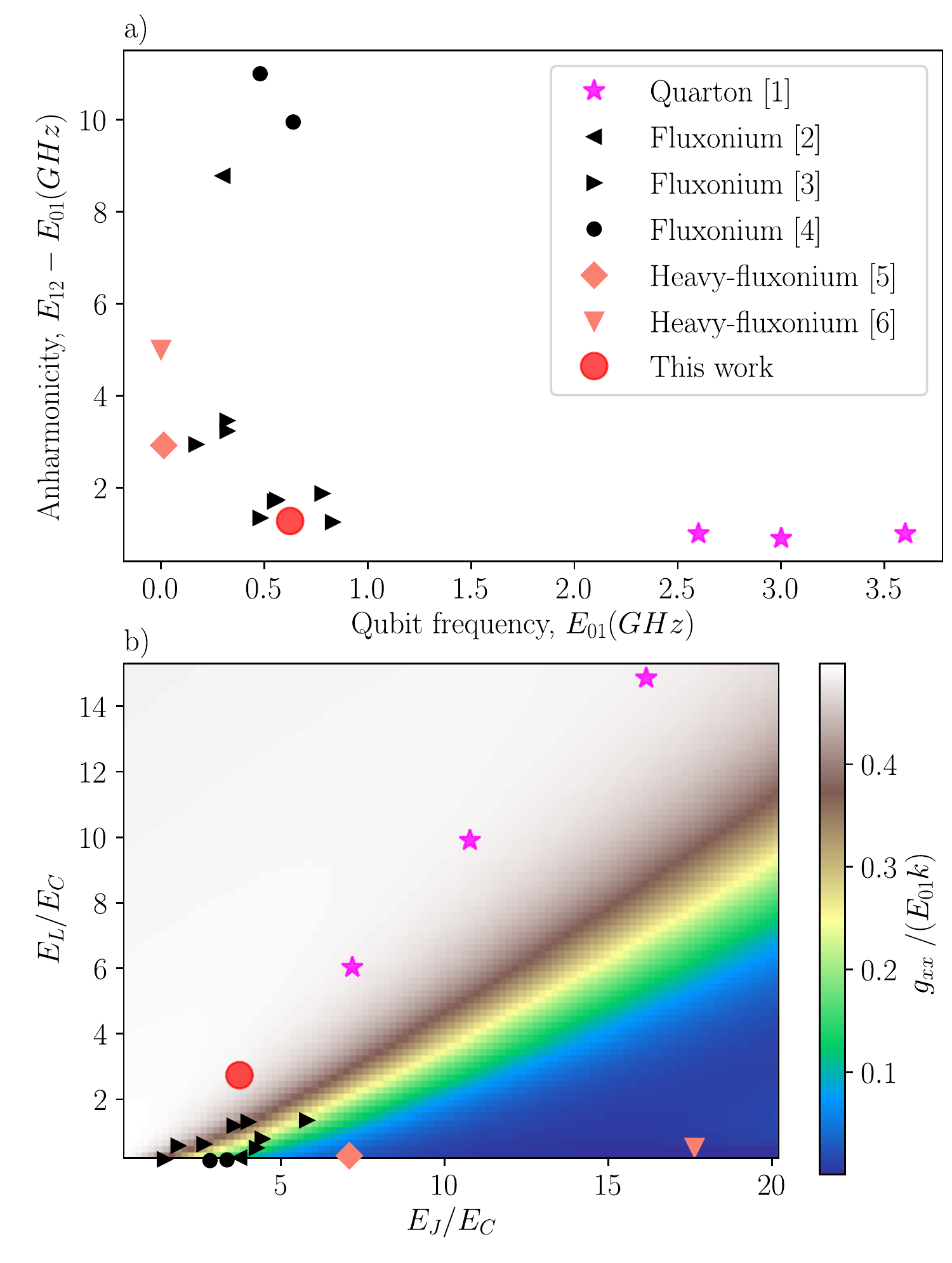}
	\caption{(color online) Dependence of the two-qubit system parameters on the qubit Josephson junction energy and inductive energy. a) Qubit frequency; b) anharmonicity; c) the effective coupling strength between two capacitively coupled qubits.}
	\label{fig:Gflux_qubit_parameters}
\end{figure}

Results of the comparison are shown in Fig.~\ref{fig:Gflux_qubit_parameters}. For presentation purposes, the parameter regimes demonstrated for regular fluxoniums \cite{Fluxonium, Nguyen, Pop}, heavy fluxoniums \cite{heavy-fluxonium, heavy}, and quarton qubits \cite{G-flux}, as well as our proposed design, are shown with solid markers. 

The frequency and coupling ratio-normalized capacitive coupling strength shown in Fig.~\ref{fig:Gflux_qubit_parameters}b is limited to $0.5$. This maximal normalized coupling is realized in the harmonic oscillator limit $E_L \gg E_J$ and in transmon qubits. We propose to operate capacitively coupled fluxoniums in the frequency regime typical for regular fluxoniums $\sim\SI{0.5}{\giga\hertz}$, while maintaining a $E_J/E_L$ ratio close to unity characteristic to quarton qubits, which does not significanlty degrade coupling strength. At the same time, the relative qubit anharmonicity is significantly larger than the asympototical 0.3 value of $E_J \gg E_C$ quartons.

It should be noted that the coupling strength degradation only applies to the fundamental qubit transition. Capacitive coupling to other transitions of fluxoniums can be effective even for $E_J/E_L\sim 10$, allowing fast two-qubit gates as shown in the work \cite{CZfluxoniums}.

\section{FULL-CIRCUIT HAMILTONIAN AND QUANTIZATION}\label{200}

\begin{figure}
	\includegraphics[width=1\columnwidth]{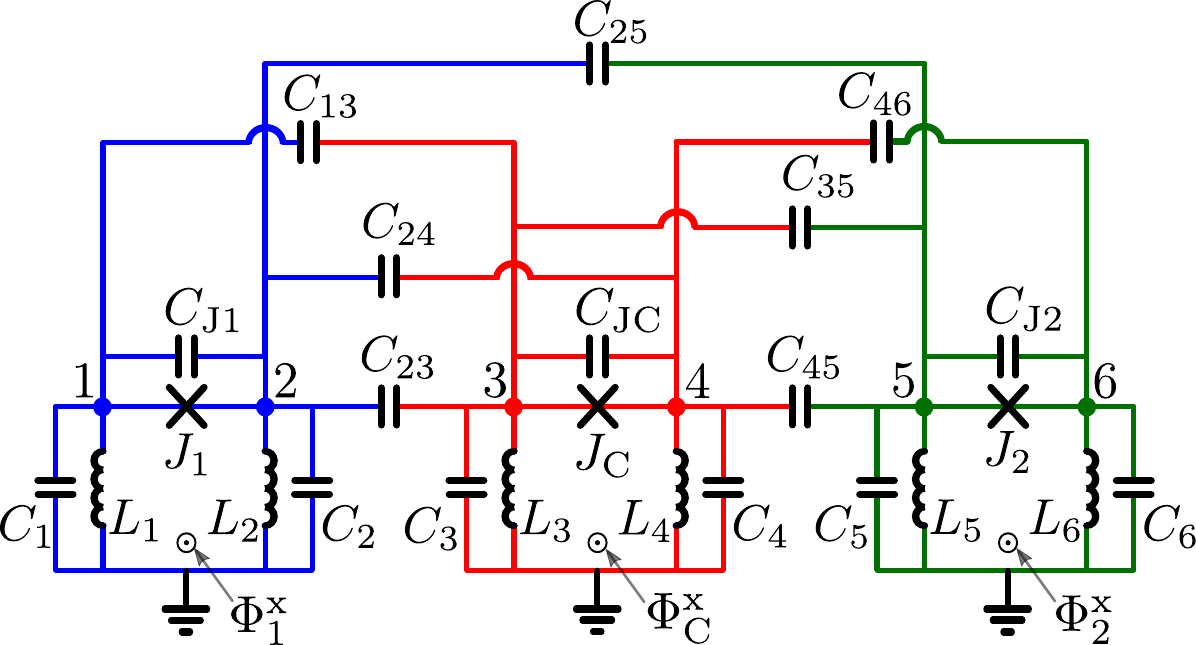}
	\caption{(color online) Equivalent lumped-element circuit for the proposed two qubit scheme with a tunable coupler. Each heavy fluxonium circuit is highlighted in different colours (qubit 1 (blue), qubit 2 (green), and coupler C (red)). $L_i$ stand for superinductors, $C_i$ stand for the electrode capacitances with respect to the ground electrode, $C_{J\lambda}$ ($\lambda = 1,C,2$) are the capacitance of Josephson junctions, $C_{ij}$ are the mutual capactitances between nodes $i$ and $j$ that facilitate coupling between the qubits.}
	\label{fig:full_schematic}
\end{figure}

The extended circuit model implementing our proposal is shown in Fig.~\ref{fig:full_schematic}. Each of the three elements is treated as a modified heavy fluxonium formed by two capacitors $C_i$, two inductors $L_i$, where $i=1,\dots,6$, and a Josephson junction $J_\lambda$, where $\lambda=1,C,2$. The external fluxes $\Phi^\textnormal{x}_\lambda$ are applied to loops of the computational qubits and coupler.

We choose node fluxes $\phi_i$, corresponding to nodes $i$ in Fig.~\ref{fig:full_schematic}, as the generalized coordinates of the system. We can write down the circuit Lagrangian $L(\phi_i,\dot{\phi_i})$ using node fluxes together with the voltages $\dot{\phi}_i$:

\begin{multline}
L = T - U, \ \ \ \ \ \ \ \ \ \ \ \ \ \ \ \ \ \
\label{eqA1}
\end{multline}
\begin{multline}
T = \frac{1}{2}\big[C_1\dot{\phi}_1^2 + C_2\dot{\phi}_2^2 +  C_\textnormal{J1}(\dot{\phi}_2-\dot{\phi}_1)^2 +  C_3\dot{\phi}_3^2 + \\ C_4\dot{\phi}_4^2 + C_\textnormal{JC}(\dot{\phi}_4-\dot{\phi}_3)^2 + C_5\dot{\phi}_5^2 +  C_6\dot{\phi}_6^2 + \\ C_\textnormal{J2}(\dot{\phi}_6-\dot{\phi}_5)^2+ 
C_{13}(\dot{\phi}_3-\dot{\phi}_1)^2 +  C_{23}(\dot{\phi}_3-\dot{\phi}_2)^2 + \\
C_{45}(\dot{\phi}_5-\dot{\phi}_4)^2 + C_{46}(\dot{\phi}_6-\dot{\phi}_4)^2 + C_{24}(\dot{\phi}_4-\dot{\phi}_2)^2 + \\ C_{35}(\dot{\phi}_5-\dot{\phi}_3)^2
C_{25}(\dot{\phi}_5-\dot{\phi}_2)^2\big],
\label{eqA2}
\end{multline}
\begin{multline}
U = E_\textnormal{J1}[1-\cos(\frac{2\pi(\phi_2-\phi_1)}{\Phi_0})] + \\ E_\textnormal{JC}[1-\cos(\frac{2\pi(\phi_4-\phi_3)}{\Phi_0})] +
E_\textnormal{J2}[1-\cos(\frac{2\pi(\phi_6-\phi_5)}{\Phi_0})] + \\ \frac{1}{2L_1}\phi_1^2 + \frac{1}{2L_2}(\phi_2 - \phi^\textnormal{x}_1)^2 + \frac{1}{2L_3}\phi_3^2 + \\ \frac{1}{2L_4}(\phi_4 - \phi^\textnormal{x}_C)^2 + \frac{1}{2L_5}\phi_5^2 + \frac{1}{2L_6}(\phi_6 - \phi^\textnormal{x}_2)^2,
\label{eqA3}
\end{multline}

where $T$ and $U$ are, respectively, the kinetic and potential energy. 

The kinetic energy term can be rewritten in matrix form $T=\frac{1}{2}\vec{\dot{\phi}}^T C_\textnormal{mat} \vec{\dot{\phi}}$, where $\vec{\dot{\phi}}=[\dot{\phi}_1, \dot{\phi}_2, \dot{\phi}_3, \dot{\phi}_4, \dot{\phi}_5, \dot{\phi}_6]$ and $C_\textnormal{mat}$ is a $6\times6$ capacitance matrix:
\begin{multline}
C_\textnormal{mat}=\begin{bmatrix}
C_\textnormal{f1}&-C_{\textnormal{J1}}&-C_{13}&0&0&0\\
-C_{\textnormal{J1}}&C_\textnormal{f2}&-C_{23}&-C_{24}&-C_{25}&0\\
-C_{13}&-C_{23}&C_\textnormal{f3}&-C_{\textnormal{JC}}&-C_{35}&0\\
0&-C_{24}&-C_{\textnormal{JC}}&C_\textnormal{f4}&-C_{45}&-C_{46}\\
0&-C_{25}&-C_{35}&-C_{45}&C_\textnormal{f5}&-C_{\textnormal{J2}}\\
0&0&0&-C_{46}&-C_{\textnormal{J2}}&C_\textnormal{f6}
\end{bmatrix}, \\
\label{eqA4}
\end{multline}
where
\begin{multline}
C_\textnormal{f1} = C_1+C_{\textnormal{J1}}+C_{13}, \ \ \ \ \ \ \ \ \ \ \ \ \ \ \ \ \\
C_\textnormal{f2} = C_2+C_{\textnormal{J1}}+C_{23}+C_{24}+C_{25}, \ \ \ \ \ \ \ \ \ \ \ \ \ \ \ \ \ \ \ \ \ \\
C_\textnormal{f3} = C_3+C_{\textnormal{JC}}+C_{13}+C_{23}+C_{35}, \ \ \ \ \ \ \ \ \ \ \ \ \ \ \ \ \ \ \ \ \ \\
C_\textnormal{f4} = C_4+C_{\textnormal{JC}}+C_{24}+C_{45}+C_{46}, \ \ \ \ \ \ \ \ \ \ \ \ \ \ \ \ \ \ \ \ \ \\
C_\textnormal{f5} = C_5+C_{\textnormal{J2}}+C_{45}+C_{35}+C_{25}, \ \ \ \ \ \ \ \ \ \ \ \ \ \ \ \ \ \ \ \ \ \\
C_\textnormal{f6} = C_6+C_{\textnormal{J2}}+C_{46}. \ \ \ \ \ \ \ \ \ \ \ \ \ \ \ \ \ \ \ \ \ \ \ \ \ \ \ \ \ \ \ \
\label{eqA5}
\end{multline}

To simplify further calculations, the superinductances and capacitances in each fluxonium are set equal, $L_1=L_2=L_{\textnormal{Q1}}$, $L_3=L_4=L_{\textnormal{QC}}$, $L_5=L_6=L_{\textnormal{Q2}}$, $C_{f1}=C_{f2}=C_{\textnormal{Q1}}$, $C_{f3}=C_{f4}=C_{\textnormal{QC}}$, $C_{f5}=C_{f6}=C_{\textnormal{Q2}}$. 

Neglecting capacitive interactions between the qubits, the circuit normal modes can be defined as
\begin{multline}
\theta^+_1 = \phi_1 + \phi_2; \ \ \ \ \ \theta^-_1 = \phi_1 - \phi_2; \ \ \ \  \\
\theta^+_C = \phi_3 + \phi_4; \ \ \ \ \ \theta^-_C = \phi_3 - \phi_4; \ \ \ \ \ \ \ \ \ \ \ \ \ \ \ \ \ \ \ \ \ \ \ \ \  \\ 
\theta^+_2 = \phi_5 + \phi_6; \ \ \ \ \ \theta^-_2 = \phi_5 - \phi_6. \ \ \ \ \ \ \ \ \ \ \ \ \ \ \ \ \ \ \   
\label{eqA6}
\end{multline}

Appling this coordinate transformation to the capacitance matrix yields
\begin{multline}
C_\textnormal{new}=T_r^T \times C_\textnormal{mat} \times T_r,  
\ \ \ \ \ \ \ \ \ \ \ \ \ \ \ \ \ \ \ \ \ 
\label{eqA7}
\end{multline}
where the transformation matrix $T_r$ is defined as:
\begin{multline}
T_r=\frac{1}{2}\begin{bmatrix}
1&1&0&0&0&0\\
1&-1&0&0&0&0\\
0&0&1&1&0&0\\
0&0&1&-1&0&0\\
0&0&0&0&1&1&\\
0&0&0&0&1&-1&
\end{bmatrix}.
\ \ \ \ \ \ \ \ \ \ \ \ \ \ \  
\label{eqA8}
\end{multline}

The potential energy becomes
\begin{multline}
U = \sum_{i=1,C,2}^{} \bigg[ E_{\textnormal{J}i}[1-\cos(\frac{2\pi\theta^-_i}{\Phi_0})] + \\ \frac{1}{4L_{\textnormal{Qi}}}(\theta^+_i - \phi^\textnormal{x}_i)^2 + \frac{1}{4L_{\textnormal{Q}i}}(\theta^-_i - \phi^\textnormal{x}_i)^2 \bigg].
\label{eqA9}
\end{multline}

We define the canonically conjugate momenta ${q^{\pm}}_i$ corresponding to the variables introduced in Eq.~\eqref{eqA6} as
\begin{multline}
q^{\pm}_i = \frac{\partial  L}{\partial {\dot{\theta}^{\pm}}_i}, \ \  \ \ \ \ \ \ \ \ \ \ \ \ \ \  \
\label{eqA10}
\end{multline}
and the canonical momentum vector $\vec{q}=[q^{+}_1, q^{-}_1, q^{+}_C, q^{-}_C, q^{+}_2, q^{-}_2]$.

The system Hamiltonian in terms of the first-order normal modes is defined as
\begin{multline}
H = \sum_{i, \alpha}^{} q^{\alpha}_i {\dot{\theta}^{\alpha}}_i - L =\frac{1}{2} \vec{q}^T C^{-1}_\textnormal{new} \vec{q} + U, \ \ \ \ \
\label{eqA11}
\end{multline}
where $C^{-1}_\textnormal{new}$ is the inverse capacitance matrix.

Finally, promoting classical degrees of freedom to quantum operators, we obtain

\begin{multline}
\hat{H} = \sum_{\alpha}^{} \hat{H}_{\alpha} +\sum_{\alpha\not= \beta}^{} \hat{H}_{\alpha \beta},
\ \
\{ \alpha,\beta \} \in \{\textnormal{h}_1,\textnormal{f}_1,\textnormal{h}_\textnormal{C},\textnormal{f}_\textnormal{C},\textnormal{h}_2,\textnormal{f}_2 \}.
\label{eqA13}
\end{multline}
The indeces $\textnormal{h}_i$ and $\textnormal{f}_j$ correspond to the Hamiltonian terms associated with the symmetric $\theta^+_i$ and antisymmetric $\theta^-_i$ mode coordinates. The symmetric modes are described by harmonic oscillator-type Hamiltonians
\begin{multline}
\hat{H}_{\textnormal{h}i} = 4E_{\textnormal{C}{\textnormal{h}i}}({\hat{n}^{+}}_i)^2 + \frac{1}{2}E_{L{\textnormal{h}i}}(\vartheta^+_i - \varphi^\textnormal{x}_i)^2, 
\label{eqA14}
\end{multline}
while the antisymmetric modes are described by fluxonium-type Hamiltonians
\begin{multline}
\hat{H}_{\textnormal{f}i} = 4E_{\textnormal{C}{\textnormal{f}i}}({\hat{n}^{-}}_i)^2 + E_{\textnormal{J}i}[1-\cos(\vartheta^-_i)] +  \frac{1}{2}E_{\textnormal{L}{\textnormal{f}i}}(\vartheta^-_i - \varphi^\textnormal{x}_i)^2.
\label{eqA15}
\end{multline}

where the dimensionless variables for the flux ${\hat{\vartheta}^{\alpha}}_i = 2\pi{\hat{\theta}^{\alpha}}_i/\Phi_0$ and their canonically conjugate Cooper pair numbers ${\hat{n}^{\alpha}}_i = {\hat{q}^{\alpha}}_i/2e$ are introduced. The inductive and capacitive energies are defined as
\begin{multline}
E_{L{\textnormal{h}i}} =E_{L{\textnormal{f}i}} =\frac{[\Phi_0/(2\pi)]^2}{2L_{\textnormal{Q}i}}, \ \ \ \ \ \ \ \ \ \ \ \ \ \ \ \ \ \ 
\end{multline}
\begin{multline}
E_{C\alpha} = \frac{e^2}{2} \left(C_{\text{new}}^{-1}\right)_{\alpha \alpha} =\frac{[\Phi_0/(2\pi)]^2}{2L_{\textnormal{Q}i}}, \ \ \ \ \ \ \ \ \ \ \ \ \ \ \ \ \ \
\label{eqA17}
\end{multline}
where $\left(C_{\text{new}}^{-1}\right)_{\alpha \alpha}$ is the diagonal matrix element of the inverse capacitance matrix corresponding to the variable $\alpha$, $\alpha \in \{\text{h}_1, \text{f}_1, \text{h}_C, \text{f}_C, \text{h}_2, \text{f}_2\}$
and the dimensionless external fluxes are defined as
\begin{multline}
\varphi^\textnormal{x}_i = \frac{2\pi}{\Phi_0}\phi^\textnormal{x}_i. \ \ \ \ \ \ \ \ \ \ \ \  \ \ \ \ \ 
\label{eqA18}
\end{multline}

The double-indexed terms $\hat{H}_{\alpha\beta}$ in Eq.\eqref{eqA13} describe the capacitive coupling between different modes. In a symmetric circuit, direct interaction between the harmonic and fluxonium modes on the same node vanish:
\begin{multline}
\hat{H}_{\textnormal{h1}\textnormal{f1}} = 0, \ \ \ \hat{H}_{\textnormal{hc}\textnormal{fc}} = 0, \ \ \  \hat{H}_{\textnormal{h2}\textnormal{f2}} = 0. \ \ \
\label{eqA20}
\end{multline}

The simplified Hamiltonian in the main text of the article Eq. 5 can be obtained by dropping the harmonic mode terms of the computational qubits, yielding
\begin{multline}
\hat{H}\subtxt{full} =  \hat{H}_{\textnormal{f1}} + \hat{H}_{\textnormal{hc}} +  \hat{H}_{\textnormal{fc}} + \hat{H}_{\textnormal{f2}} + \hat{H}\subtxt{V}, \ \ \ \ \ 
\label{eqA21}
\end{multline}
where the interaction $\hat{H}\subtxt{V}$ of two qubits consists of five terms: the direct coupling ($\hat{H}_{\textnormal{f1}\textnormal{f2}}$), the indirect coupling via the coupler harmonic mode ($\hat{H}_{\textnormal{f1}\textnormal{hc}}$ and $\hat{H}_{\textnormal{hc}\textnormal{f2}}$) and the indirect coupling via the coupler fluxonium mode ($\hat{H}_{\textnormal{f1}\textnormal{fc}}$ and $\hat{H}_{\textnormal{fc}\textnormal{f2}}$).

Note that this description is not entirely accurate, as the harmonic modes do interact with the fluxonium modes of the computational qubit due to their coupling to the coupler's modes. Moreover, circuit asymmetry and nonlinearity in the superinductor can also contribute to the interaction between the fluxonium and harmonic modes on a single node. The contribution of the harmonic modes of the qubits to the effective qubit-qubit interactions leads to a small renormalization of the low-energy Hamiltonian. We include these modes in our static Hamiltonian simulations, specifically for the static ZZ-interaction, and neglect them in the gate simulations.

The circuit parameters used for the following calculations are $C_{1} = C_{6} = 70.53\ \si{\femto\farad}$, $C_{2} = C_{5} = 51.17\ \si{\femto\farad}$, $C_{3} = C_{4} = 49.17\ \si{\femto\farad}$, $C_{J1} = C_{JC} =C_{J2}= 1.056\ \si{\femto\farad}$, $C_{25} = 0.167\ \si{\femto\farad}$, $C_{23} = C_{45} = 19.20\ \si{\femto\farad}$, $C_{13} = C_{46} = 0.176\ \si{\femto\farad}$, $C_{24} = C_{35} = 0.234\ \si{\femto\farad}$, $E_\textnormal{J1} = E_\textnormal{JC} = E_\textnormal{J2} = 2.14\  \si{\giga\hertz}$, $E_{L1} = E_{L2} = E_{L5} = E_{L6} = 1.514\  \si{\giga\hertz}$, $E_{L3} = E_{L4} = 1.634\  \si{\giga\hertz}$. This choice of capacitances allowed us to reach the desired values of qubit frequencies and effective qubit-qubit coupling. The Josephson junction energies and inductive energies are accessible within the fabrication techniques used in our previous work \cite{Planar_fluxonium}. For the phase slip element we propose to use a $S_1\approx100\times90\  \si{\nano\meter}^2$ Josephson junction, and for the superinductance an array ($N\approx80$) of series-connected of big Josephson junctions ($S_2\approx1000\times500\  \si{\nano\meter}^2$). All junctions can be fabricated by the shadow evaporation technique with critical current density $j=0.5\  \si{\micro\A}/\si{\micro\meter}^2$.

\section{NUMERICAL RESULTS}\label{300}

In this Appendix we present the results of numerical calculation of the full system Hamiltonian. We found the eigenvalues and charge matrix elements for all independent fluxonium and harmonic modes from Eqs.~\eqref{eqA14},\eqref{eqA15} using numerical diagonalization.
The data qubits are design to be kept in the lower flux sweet spot ($\varphi_{1,2}^\text{x} = \pi$), while the magnetic flux in the coupler loop is varied between zero flux and half flux quantum ($\varphi_C^\text{x} \in \left[0, \pi\right]$). 

\begin{figure}
	\includegraphics[width=1\columnwidth]{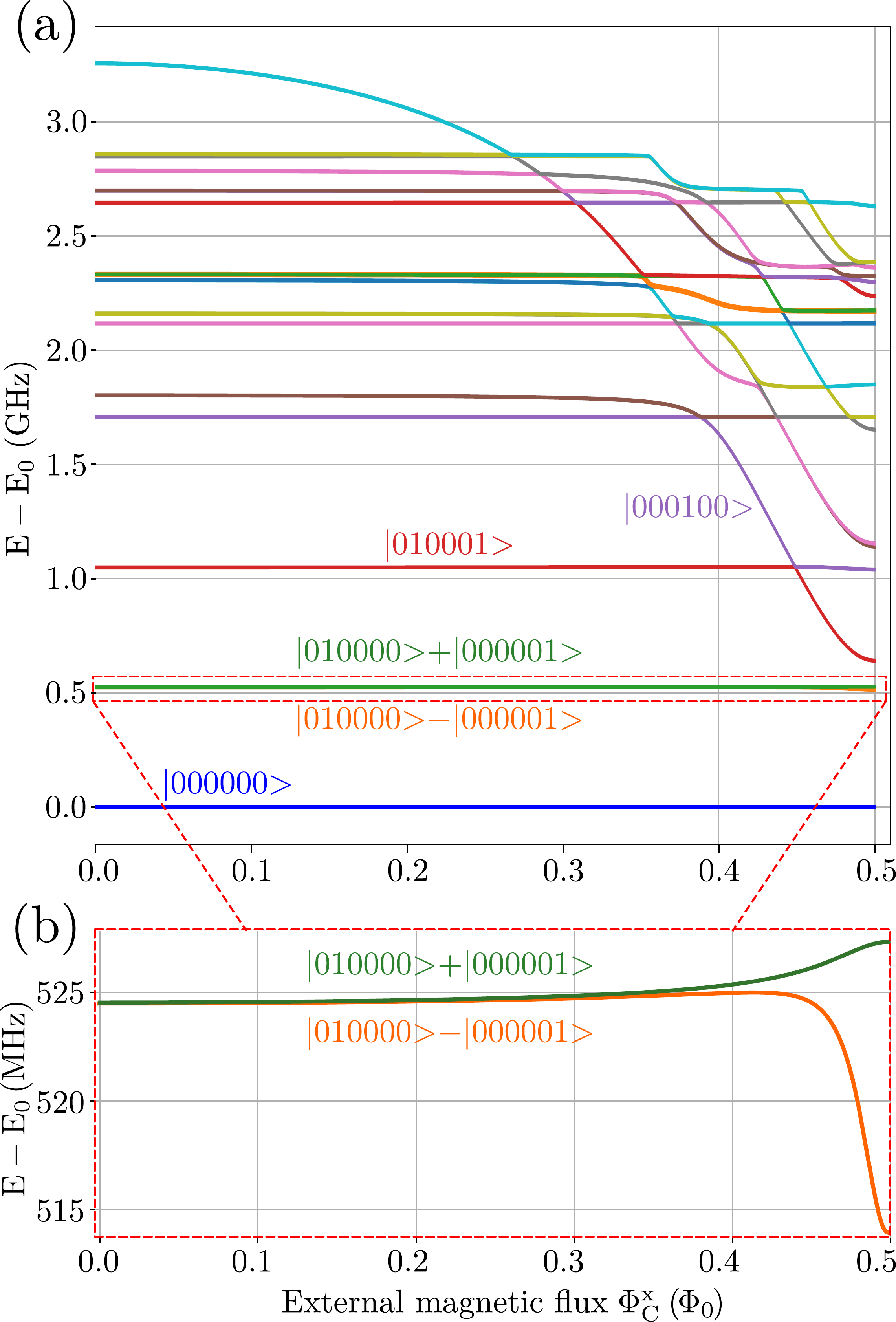}
	\caption{(color online) a) Energy levels of the tunable system vs magnetic flux in the coupler ${\Phi}^{\textnormal{x}}_\textnormal{C}$. b) The red dotted rectangle outlines eigenenergies of the data qubits one-excitation manifold.}
	\label{fig:full_spectr}
\end{figure}

To specify the complete Hamiltonian we used the open-source QuTiP\cite{QuTiP} package. In each fluxonium-type mode we took the first five levels, and in each harmonic mode we took the first three levels and used corresponding matrix elements to take into account the terms responsible for the interaction \eqref{eqA21}. Finally, we numerically diagonalized the full Hamiltonian. The computed energy spectrum as a function of magnetic flux ${\Phi}^{\textnormal{x}}_\textnormal{C}$ is plotted in Fig.~\ref{fig:full_spectr}a. 

Full system eigenstates are labeled as $\ket{ n\subtxt{h1}, n\subtxt{f1}, n\subtxt{hc}, n\subtxt{fc},n\subtxt{h2}, n\subtxt{f2} }$, 
where $n_{\alpha}$ is 
the occupancy of the $\alpha$-mode, $\alpha \in \{\text{h}_1, \text{f}_1, \text{h}_C, \text{f}_C, \text{h}_2, \text{f}_2\}$.
The five lowest-lying levels are labeled in Fig.~\ref{fig:full_spectr}a. These levels play a key role in the two-qubit gates. Since the computational levels of first qubit $\ket{010000}$ and second qubit $\ket{000001}$ are degenerate (Fig.~\ref{fig:full_spectr}b), the eigenstates are their symmetric (green line) and antisymmetric (orange line) combinations, and the energy gap between these states corresponds to the effective $XX$ coupling.

\begin{figure*}
	\includegraphics[width=2\columnwidth]{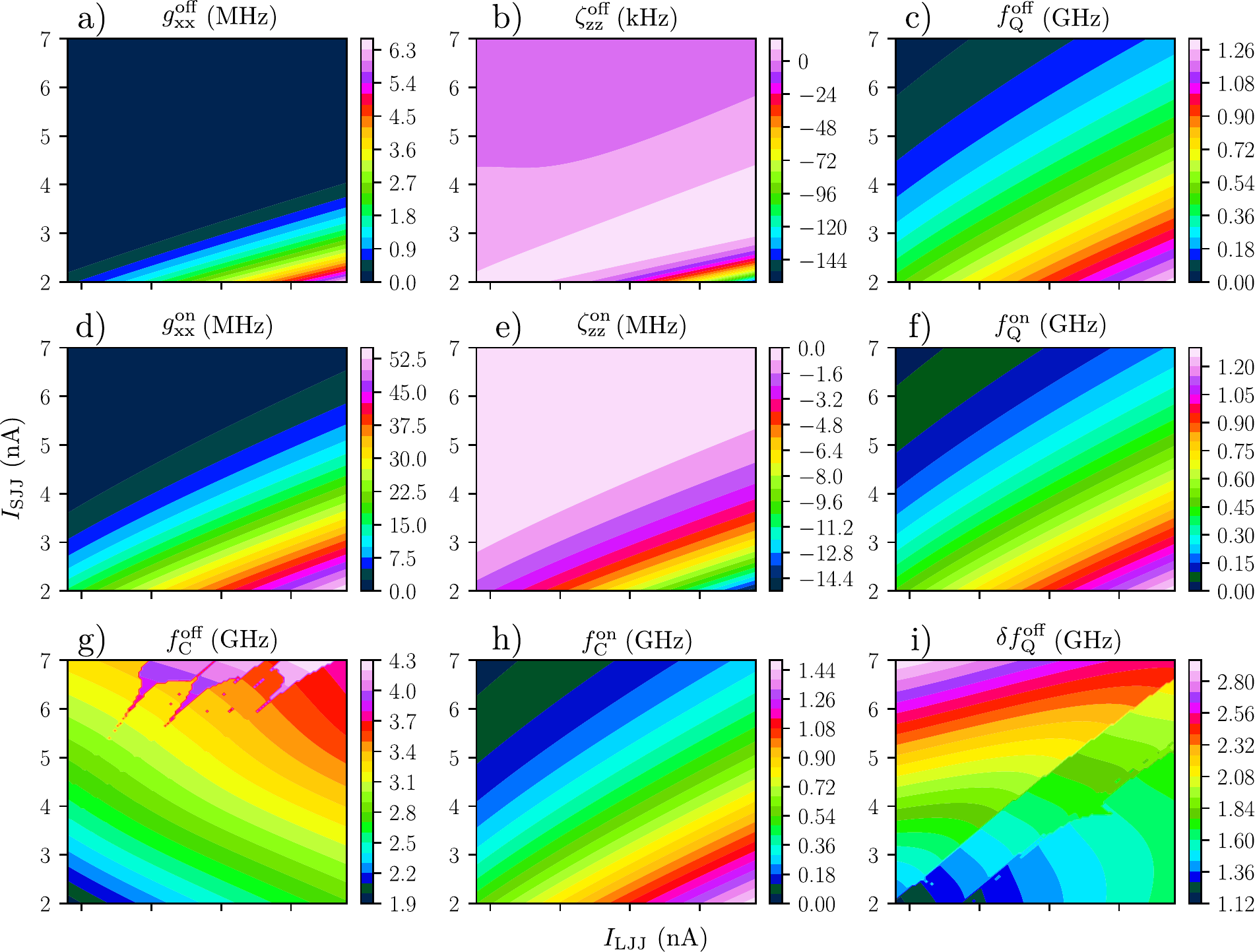}
	\caption{(color online) Dependence of the low-energy effective Hamiltonian parameters on the critical current of small and large Josephson junctions. a) The effective coupling at the zero flux bias point $g^\textnormal{off}_\textnormal{xx} = g_\textnormal{xx}(\Phi_\textnormal{C} = 0)$; b) the effective coupling at the flux degeneracy point $g^\textnormal{on}_\textnormal{xx} = g_\textnormal{xx}(\Phi_\textnormal{C} = \Phi_0)$; d) parasitic ZZ crosstalk at the zero flux bias point $\zeta^\textnormal{off}_\textnormal{zz}$; e) parasitic ZZ crosstalk at the flux degeneracy point $\zeta^\textnormal{on}_\textnormal{zz}$. c),f),g),h) Qubit and coupler frequencies $f^\textnormal{off}_\textnormal{Q}$ and $f^\textnormal{on}_\textnormal{Q}$,  $f^\textnormal{off}_\textnormal{C}$ and $f^\textnormal{on}_\textnormal{C}$ at the zero flux bias point and at the flux degeneracy point of the coupler. i) Data qubit anharmonicity $\delta f^\textnormal{off}_\textnormal{Q}$.}
	\label{fig:Parameters}
\end{figure*}

\section{CRITICAL CURRENT DEPENDENCE}\label{400}

A crucial issue for large scale Josephson junction based circuits is robustness with respect to critical current deviations of small junctions. The aim of this section is to identify how these deviations affect the effective low-energy Hamiltonian parameters.
We sweep the critical current value of small Josephson junctions used as the nonlinear element for data qubits and coupler (for simplicity we consider them the same) and large Josephson junctions used in superinductances arrays. The data qubits' superinductances consist of 41 junctions, while the coupler's superindutances have 38 junctions each, which results in the coupler frequency being $\approx100\ \mathrm{MHz}$ higher in the flux degeneracy point.
The result of this calculation are shown in Fig.~\ref{fig:Parameters}.
\begin{figure}
	\includegraphics[width=0.8\columnwidth]{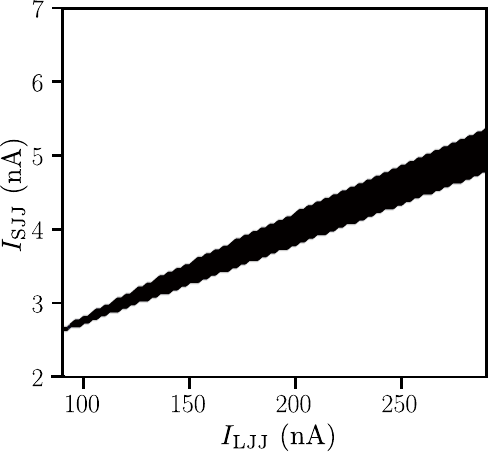}
	\label{fig:Suitable_values}
	\caption{Suitable critical current values. Black area indicates the range of critical currents values allowing one to implement the proposed scheme of two fluxonium qubits in the desired range of low energy effective Hamiltonian parameters.}
\end{figure}

Here we found the effective coupling at the zero flux bias point and the flux degeneracy point in the coupler loop ($g^\textnormal{off}_\textnormal{xx}$ and $g^\textnormal{on}_\textnormal{xx}$ respectively) as well as parasitic ZZ crosstalk ($\zeta^\textnormal{off}_\textnormal{zz}$ and $\zeta^\textnormal{on}_\textnormal{zz}$ respectively). We also defined data qubits frequencies $f^\textnormal{off}_\textnormal{Q}$ and $f^\textnormal{on}_\textnormal{Q}$ and coupler frequencies $f^\textnormal{off}_\textnormal{C}$ and $f^\textnormal{on}_\textnormal{C}$ at the coupler zero flux bias point and the flux degeneracy point. For the sake of completeness we also present here data qubit anharmonicity $\delta f^\textnormal{off}_\textnormal{Q}$.
Fig.~\ref{fig:Suitable_values} shows the region (black area) with suitable critical current values, at which the proposed tunable coupling scheme can be physically implemented. This region was defined from the conditions: $8\   \mathrm{MHz}<g^\textnormal{on}_\textnormal{xx}< 30\  \mathrm{MHz}$, $g^\textnormal{off}_\textnormal{xx}< 0.5\  \mathrm{MHz}$, $|\zeta^\textnormal{off}_\textnormal{zz}|<5\  \mathrm{kHz}$, $|\zeta^\textnormal{on}_\textnormal{zz}|<1.5\  \mathrm{MHz}$, $200\  \mathrm{MHz} < f^\textnormal{off}_\textnormal{Q}<600\  \mathrm{MHz}$, $\delta f^\textnormal{off}_\textnormal{Q}>1.2\  \mathrm{GHz}$. It should be noted that the Fig.~\ref{fig:Suitable_values} is shown as an example and the selected conditions are not strict.

\section{CONSTRUCTION OF THE CPHASE GATE}\label{500}

The control parameter used to implement the two-qubit gates, the coupler flux, changes both qubit frequencies, XX and ZZ couplings at the same time. As a result, the two-qubit gate family that can be implemented using this method is equivalent to $\operatorname{fSim}(\theta, \varphi)$, with both $\theta$ and $\phi$ somehow depending on the control signal $\Phi_C^x(t)$ applied to the coupler flux line. 

\begin{figure*}
	\begin{quantikz}
		\qw & \gate[]{H} & \gate[]{U_1(-\varphi)} &\gate[2]{\mathrm{fSim}(\frac{\pi}{4}, \varphi)} &  \gate[]{X} & \gate[2]{\mathrm{fSim}(\frac{\pi}{4}, \varphi)} & \gate[]{U_1(-\varphi)} & \gate[]{H} & \gate[]{S} &\qw\\
		\qw & \gate[]{H} & \gate[]{U_1(-\varphi)} &  & \qw & & \gate[]{U_1(-\varphi)} &\gate[]{H} & \gate[]{S^\dagger}& \qw 
	\end{quantikz}
	\caption{Construction of the CPHASE gate from two fSim gates with $\theta=\pi/4$ and arbitrary conditional phase angle $\varphi$}. 
	\label{fig:cphase}
\end{figure*}
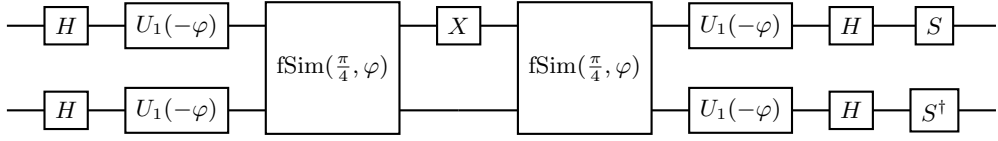

A wide range of quantum algorithms relies on the CPHASE gate. To construct the CPHASE we gate using our proposed two-qubit scheme, we propose the spin-echo technique initially devised to remove the conditional phase from cross-resonance gates \cite{Corcoles2013}. The gate sequence implementing a CPHASE gate is shown in Fig.~\ref{fig:cphase}. The gate sequence consists of two two-qubit fSim gates interleaved by single-qubit gates. In applications the single-qubit gates before and after the fSim gates can be merged together with other gates for better fidelity.

\section{COUPLING OF HARMONIC AND FLUXONIUM MODE}\label{600}
The presence of finite asymmetry in the qubit capacitances and inductances translates into coupling between the harmonic and fluxonium mode. For a single qubit circuit, we introduce  the capacitance and inductance asymmetries $\delta C, \delta L$, with $C_1 = C + \delta C / 2$, $C_2 = C - \delta C / 2$, $L_1 = L + \delta L / 2$, $L_2 = L - \delta L / 2$.

For small asymmetries the Hamiltonian perturbation is defined as
\begin{multline}
\hat{V} = \frac{4e^2\delta C}{C(C+2C_J) - \delta C^2/4} \hat{n}_f\hat{n}_h + \\
\frac{\hbar^2}{4e^2}\frac{\delta L}{L^2 - \delta L^2/4} \hat{\varphi}_f\hat{\varphi}_h,
\end{multline}
which is Jaynes-Cummings type Hamiltonian for a fluxonium qubit coupled to a resonator. When the qubit is biased at half flux quantum, the frequency detuning between the two modes is large. In this dispersive regime excitations in the resonator mode induce a dispersive shift $\chi$ in the qubit frequency which is quadratic in $\delta L$ and $\delta C$.
For relative asymmetries $\delta C/C$ and $\delta L/L$ of 5\% in both capacitance and inductance the dispersive shift arising from this coupling is $\chi = 23~\mathrm{MHz}$.

Another source of dispersive shifts is the nonlinearilty of the superinductors. In the proposed design with $N=80$ junction in each inductor, the first nonlinear correction to the superinductance Hamiltonian is given by
\begin{equation}
\hat{V} = - \frac{E_L}{24 N^2}\left(\hat{\varphi}_f + \hat{\varphi}_h\right)^4.
\end{equation}

From first-order perturbation theory we obtain a cross-Kerr coefficient of $\chi=0.5~\mathrm{MHz}$.

Similar to the case of 0-$\pi$ qubits, thermal excitation may degrade qubit coherence times \cite{Groszkowski2018}. The pure dephasing rate associated with this process can be estimated for low thermal harmonic mode occupancies $n_\mathrm{th}$ by the formula\cite{Wang2019}
\begin{equation}
\gamma_\varphi = \frac{n_\mathrm{th}\kappa \chi^2}{\chi^2+\kappa^2},
\label{thermal}
\end{equation}
where $n_\mathrm{th}$ is the photon number, $\kappa$ is the harmonic mode decay rate and $\chi$ is the dispersive shift. We expect that in real devices $\chi \gg \kappa$. The thermal population of the harmonic mode can be estimated as $n_{\mathrm{th}} \approx 10^{-4}$ for $T=10~\mathrm{mK}$.

The decay rate for the harmonic mode $\kappa$ can be obtained through Fermi's Golden rule:
\begin{equation}
\kappa = \omega  \frac{2\pi Z_0}{R_Q}\left(\frac{C_{\mathrm{an}}}{C+C_{\mathrm{an}}}\right)^2 \left|\langle 0 |\hat{n}^{+}|1\rangle\right|^2,
\end{equation}

where $\omega$ is the harmonic mode frequency, $Z_0 = \SI{50}{\ohm}$ is the control line impedance, $R_Q$ is the von Klitzing constant, and $\langle 0 |\hat{n}^{+}|1\rangle$ is the matrix element of the harmonic mode charge operator for the fundamental transition.
We choose $C_{\mathrm{an}}=0.34~\mathrm{fF}$ for the coupling capacitance with microwave antenna (Fig.~1 from the main article), which corresponds to a decay rate $\kappa=0.01~\mathrm{MHz}$. From Eq.~\eqref{thermal} we obtain $T_{\varphi}>1~\mathrm{s}$ for $T=10~\mathrm{mK}$.


\end{document}